\documentclass[doublecol]{epl2}
\usepackage{amsmath}
\usepackage{graphicx}
\usepackage{amssymb}
\usepackage{latexsym}
\usepackage[colorlinks,bookmarksopen]{hyperref}

\newcommand\rs[1]{{\scriptscriptstyle\rm #1}}

\title{Optical stabilization of voltage fluctuations in half-Josephson lasers}
\author{Frans Godschalk \and Yuli V. Nazarov}
\institute{Kavli Institute of Nanoscience - Delft University of Technology,
P.O.  Box 5046, 2600 GA Delft, The Netherlands}


\abstract{
A recently proposed device, dubbed half-Josephson laser, provides a phase-lock
between the optical phase and the superconducting phase difference between the
leads of the device. In this paper we propose to utilize this phase-lock for
stabilization of voltage fluctuations, by two optical feedback schemes.
The first scheme involves a single half-Josephson laser and allows to
significantly decrease the diffusion coefficient of the superconducting phase
difference. The second scheme involves a stable optical source and a fluctuating
half-Josephson laser and permits quenching of the diffusion of the relative
phase of the lasers. This opens up perspectives of the optical control of the
superconducting phase and voltage fluctuations.
}

\begin{document}

\maketitle
In the past years technological developments have led to the realization of
novel hybrid devices that combine superconductors and
semiconductors\cite{DeFranceschi}. These hybrid devices form a basis to study
materials\cite{hayat}, realize new functionalities, like in a supercurrent
transistor\cite{katsaros}, and investigate exotic states of matter, like
topological superconductivity\cite{mourik}. As a part of this development,
superconducting optoelectronic devices have been realized and proposed,
where the interaction between light and superconductor-semiconductor
structures is important. Enhanced emission of light-emitting diodes (LED) has
been demonstrated\cite{sasakura}. Interesting theoretical proposals include the
Josephson LED\cite{recher}, the half-Josephson laser\cite{godschalk,
godschalk2} (HJL) and devices useful for quantum information
purposes\cite{Khoshnegar}. 
 
In this letter, we explore some intriguing applications which exploit the most
important property of the HJL: a phase-lock between optical
phase and superconducting phase difference. The HJL consists of a
biased Josephson junction containing a structure capable of emitting light by
electron-hole recombination. The light is emitted in an optical resonator.
Importantly, the eigenstates associated with the light emission also couples
to the superconducting leads. As a result of this, the optical phase is proven
to be locked to the superconducting phase difference. In ref.~\cite{godschalk},
a HJL model based on a single quantum emitter was studied, whereas a
model for many quantum emitters was considered in ref.~\cite{godschalk2}. In the
latter, the coherence time of the optical phase was found to be exponentially
long. Decoherence is caused only by switchings between stable states of
radiation, corresponding to two locked values of the optical phase. The
phase-lock thus quenches the phase diffusion inevitable in usual lasers.

In previous studies, the superconducting phase difference was assumed to
be fixed. For any physical realization of the HJL, however, the
superconducting phase difference is expected to diffuse. The diffusion
coefficient is proportional to the zero frequency noise of the bias voltage.
As a consequence of the phase lock, also the optical phase will be
subjected to this diffusion, thus limiting the coherence time of the Josephson
laser. 
In this letter, we describe two optical feedback schemes, that exploit the
phase-lock in the HJL, to stabilize the fluctuations of the bias
voltage. The first scheme involves a single HJL and significantly
decreases the diffusion coefficient of the superconducting phase
difference. The second scheme involves the optical locking of a fluctuating HJL
to a stable optical source. This can be exploited to control the
superconducting phase of the HJL and create voltage pulses, by changing the
optical phase in time. The optical phase is changed by changing the optical path
lengths of the laser beams.

\section{The half-Josephson laser}
We start with a brief overview of the dynamics of the HJL, as it is described in
\cite{godschalk2} in more detail, and we will specify the feedback scheme later.

The HJL is driven by a dipole moment that oscillates with half the Josephson
frequency, $\omega_J/2$, corresponding to the average bias voltage, and is
composed of individual contributions of a large number of quantum emitters. As
in the case of any laser, the dipole moment saturates with increasing field
strength in the resonant mode. The dipole moment originates from the coupling of
the quantum emitters to the superconducting leads. The dipole moment fluctuates
due to quantum noise in the optical field, as in usual lasers \cite{scullylamb},
and also due to spontaneous
switchings of the quantum emitters between their eigenstates. The latter
fluctuations can be seen as a renormalization of the quantum noise.
 
In \cite{godschalk2}, we derive the semiclassical equations of motion for the
optical field in the HJL, which do not depend on microscopic
details, and provide the basis of the phenomenological description of the HJL.
Under assumption of weak coupling of the quantum emitters to the
superconductivity and the optical field, the equations of motion are given by
	\begin{align}\label{eq:feedbackgeneral}
	   \begin{split}
	      \frac{d}{dt}|b| &= -\frac{\Gamma}{2}|b|-A|b|\sin(2\Phi_b-\Phi_\Delta)+
					\xi_{|b|}(t),\\
			\dot\Phi_b &= -\omega -A\cos(2\Phi_b-\Phi_\Delta) -
					\Omega'' |b|^2 + \xi_\phi(t), \\ 
			\dot\Phi_\Delta &= \frac{2e}{\hbar}\tilde v(t).
	   \end{split}
	\end{align}
Here, the optical field is represented by $b=\langle \hat b\rangle$, the
expectation value of the photon annihilation operator, with phase $\Phi_b$. The
phase $\Phi_\Delta$ is the superconducting phase difference across the Josephson
junction, in a rotating frame of reference. Because of the phase lock only the
phase combination $2\Phi_b-\Phi_\Delta$ occurs at the righthandside of this
equation. In contrast to what is the case now, $\Phi_\Delta$ was taken constant
in ref.~\cite{godschalk2}. As regards the other parameters, $\omega$ is the
detuning of the photon frequency with respect tot
the resonator frequency, $A$ and $\Omega''$ are coefficients with a value
determined by the dipole moment, $\tilde v(t)$ is the time dependent fluctuation
of the voltage bias and $\Gamma$ is the decay rate of the resonator.
The quantities $\xi_{|b|}(t)$, $\xi_\phi(t)$ and $\tilde v(t)$ are Langevin
noise sources, with zero time average and satisfying
	\begin{align}
		\langle  \xi_{|b|}(t) \xi_{|b|}(t')\rangle &= 
				\frac{\Gamma}{4}\delta(t-t') = 
	   n_s \langle \xi_\phi(t)\xi_\phi(t')\rangle, \nonumber\\ 
		\langle \tilde v(t)\tilde v(t')\rangle &= k_B T Z \delta(t-t'),
	\end{align}
with $n_s$ the stationary number of photons in the resonator, $Z$ the
impedance of the junction and $k_B T$ the thermal energy.
The stationary solutions to eq.~(\ref{eq:feedbackgeneral}) describe steady state
lasing with a fixed value of the phase combination $2\Phi_b-\Phi_\Delta$. The
two phases are indeed locked. 

To study noise in the HJL, eq.~(\ref{eq:feedbackgeneral}) can be
simplified by linearizing it about its stationary value. Surely,
fluctuations of the optical field are expected to be small compared to the
field itself, in the steady state operation of the laser. Taking frequency in
units of $\Gamma/2$, the linearized equations in Fourier space are given by
	\begin{align}\label{eq:lineqsmotion}
	   C_1(\phi_\Delta^\nu-2\phi_b^\nu) - i \nu a_\nu &= \xi_{|b|}^\nu, 
			\nonumber\\
		(2-i\nu)\phi_b^\nu - \phi_\Delta^\nu + C_2  a_\nu &=
				\xi_\phi^\nu, \\
		-i\nu \phi_\Delta^\nu &= \frac{2e}{\hbar}\tilde v_\nu, \nonumber
	\end{align}
where $a$, $\phi_b$ and $\phi_\Delta$ (and their Fourier transforms) are
respectively the deviation from $|b_s|$, $\Phi_b^s$ and $\Phi_\Delta^s$, the
stationary solutions to eq.~(\ref{eq:feedbackgeneral}), and $\nu$ is
the dimensionless frequency in a rotating frame of reference. It is taken in
units of $\Gamma/2$. The coefficients are
given by $C_1 = |b_s|(\omega + \Omega''|b_s|^2)$ and $C_2 = 2|b_s|\Omega''$. 

The linearized equations can be further reduced. First, as we are only
interested in the time dependence of the phases, we eliminate the term $C_2 
a_\nu$ in the second line of eq.~(\ref{eq:lineqsmotion}), using the equation in
the first line. Second, we concentrate on a small frequency scale, $\nu\ll1$,
where $\phi_b(t)$ adiabatically follows $\phi_\Delta(t)$. Hence, eliminating
$a_\nu$ from eq.~(\ref{eq:lineqsmotion}) and assuming that the relevant
frequencies satisfy $\nu\ll1$, we arrive at
	\begin{align}\label{eq:lineqsmotion2}
		\begin{split}
		C_1 C_2[\phi_\Delta^\nu - 2\phi_b^\nu] &= i\nu \xi_\phi^\nu + 
			C_2 \xi_{|b|}^\nu , \\
		-i\nu \phi_\Delta^\nu &= \frac{2e}{\hbar}\tilde v_\nu, 
		\end{split}
	\end{align}
The first of these equations describes fluctuations of the combined phase
$\phi_\Delta-2\phi_b$, which is not subjected to drift. These
fluctuations manifest only at large frequency scales, or 
differently stated, at short timescales. Since we concentrate on 
small frequency scales, corresponding to long timescales, we can 
assume that the effect of the high frequency fluctuations on the 
phases has averaged out to zero. Therefore we assume to have the 
time averaged value of the phase difference $\phi_\Delta - 2\phi_b$.
Consequently, as required, we have $\phi_\Delta=2\phi_b$. 
The second of these equations describes diffusion of the phases. Indeed, the
variance at low frequencies satisfies $\langle|\phi_\Delta^\nu|^2\rangle \sim
|\tilde v_\nu|^2/\nu^2$. In the time domain the variances are proportional
to time. This is very  much like phase diffusion in common lasers.

\begin{figure}
  \onefigure{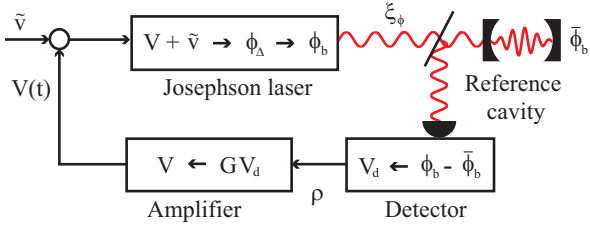}
  \caption{% 
	The feedback scheme used for stabilization of voltage fluctuations, as
described in the main text. 
The blocks represent the elements of the feedback
loop. Each element is characterized by a type of conversion of quantities, for
instance conversion of phase difference to voltage, in the detector.
The Josephon laser contains two kinds of conversions: voltage is converted
to superconducting phase difference, which in turn is converted to optical
phase. Solid (wavy) lines represent an electrical (optical) connection between
the elements. Noise sources are indicated near these lines.
  }\label{fig:feedback}
\end{figure}

\section{Feedback}
Our purpose is to stabilize the fluctuations of the bias voltage of
the HJL, by implementing a feedback loop involving the measurement of the
optical phase. Since $\phi_b$ and $\phi_\Delta$ are locked, the measuring of
the optical phase will give information about the fluctuation of
the superconducting phase. This fluctuation is then corrected by a proper
voltage feedback signal.
 
In our setup we use the well-known
{\it Pound-Drever-Hall} stabilizer~\cite{Drever1983}, which is a very
powerful scheme for stabilization of common lasers~\cite{Day1992}. Essentially,
in this scheme, the laser beam is reflected of a reference cavity with a high
quality factor, which acts as a phase memory element. The reflected beam
is then a superposition of the incident beam and a beam that leaks from the
reference cavity. Its intensity will therefore depend on the phase
difference $\phi_b-\bar\phi_b$, with $\bar\phi_b$ the time average of $\phi_b$.
The intensity is transferred to a voltage error signal, using a photo detector.
The proportionality constant is given by $\hbar\gamma/2e$, with frequency
$\gamma$.
Subsequently, the voltage error signal is amplified, with a factor $G$, and
added to the voltage noise that is already applied to the laser. The error
signal will influence the optical phase, thus closing the feedback loop. 
The total feedback signal is given by $V(t)= -G [(\hbar\gamma/2e)(\phi_b(t) -
\bar\phi_b(t)) + \varrho(t)]$, where the sign is chosen for future convenience
and $\varrho(t)$ is the noise of the amplifier. The Fourier transform of $V(t)$
is given by
	\begin{align}\label{eq:feedbackvoltage}
	   V(\nu) = -G\left[\frac{\hbar\gamma}{2e}H(\nu)\phi_b^\nu +
		\varrho_\nu\right],\, H(\nu)= 1-\frac{e^{i\nu t_a}-1}{i\nu t_a}.
	\end{align}
The time average of $\phi_b$ was taken from time $t-t_a$ to $t$, with time
$t_a$ being proportional to the quality factor of the reference cavity. We
require a reference cavity with a high quality factor, compared to that of the
HJL, implying $t_a\gg \Gamma^{-1}$. We assumed to have no time delay between
signal measurement and feedback at a timescale $t_a$~\cite{smithpredictor}.
Regarding $H(\nu)$ we
mention two important limits. For $\nu t_a\ll1$, $H(\nu) \simeq -i\nu t_a/2$,
while for $\nu t_a\gg1$, $H(\nu) \simeq 1$. Hence, in the first case, the
feedback scheme is sensitive to optical phase changes, while in the second case,
it is sensitive to the optical phase itself~\cite{Drever1983}.

Equation~(\ref{eq:lineqsmotion2}), with $\tilde v_\nu \to \tilde v_\nu +
V(\nu)$, describes noise in the HJL, with feedback at small frequencies,
$\nu\ll1$. A schematic of the feedback is given in fig.~\ref{fig:feedback}.

\section{Variance and stability}
Using the above equations, we investigate the time dependent variance of the 
optical phase. We concentrate on the variance of the difference between the
phase at time zero and time $t$
	\begin{align}\label{eq:defvariance}
		\langle[\phi_b(0)-\phi_b(t)]^2 \rangle &= 2\int
		\langle|\phi_b^\nu|^2
			\rangle \left[1-e^{-i\nu t}\right] \frac{d\nu}{2\pi}, \\
		\frac{\hbar}{2e}\phi_b^\nu &= \frac{G\varrho_\nu -
			\tilde v_\nu}{2i\nu t_a-z\ H(\nu)} t_a = \frac{\hbar}{4e}
			\phi_\Delta^\nu. \label{eq:defvariance2}
	\end{align}
Here $z\equiv G\gamma t_a$, plays the role of effective amplification
coefficient of our feedback circuit. By increasing the amplification coefficient
$G$, we can make $z$ as big as we want. We should only  make sure that the poles
of this expression lie at frequencies much smaller than $\Gamma$. We need
to require this for our approximation to be valid. We also don't expect the
feedback to work without delay, at frequencies of the order $\Gamma$. This sets
the maximum possible value of $z\simeq \Gamma t_a$. 
With this condition for z satisfied, the assumption to neglect 
the optical noise sources $\xi^\nu_\phi$ and $\xi^\nu_{|b|}$ in the 
calculation of the variance is justified.
Previously, we argued that these noise sources are irrelevant, 
since they manifest themselves at high frequencies, while it was assumed 
that only the low frequencies are relevant. The 
condition $z \lesssim \Gamma t_a$ now guarantees that this assumption 
is valid so that the high frequency fluctuations, and 
with that, also the optical noise sources, are indeed irrelevant.
 
Before calculating the variances, we need to ensure that circuit with the
feedback remains stable. The stability is governed by the positions of the
poles of the susceptibility functions, defined by eq.~\eqref{eq:defvariance2}.
Poles below (above) the real axis represent solutions that decay (grow)
exponentially with
time, while poles at the real axis represent diffusing solutions that grow
linearly with time. The relatively complex form of $H(\nu)$ prohibits us
from finding the poles positions explicitly. Instead, we
look for values of the parameter $z$, at which the poles cross the real
axis. Thereby we find the boundary of the stability region, since the circuit
is stable without feedback, which is at $z=0$.
To find this value of $z$, we need to solve
	\begin{align}\label{eq:pole}
	   2i\nu t_a-z\left[1- \frac{e^{i\nu t_a}-1}{i\nu t_a} \right] = 0,
	\end{align}
for real $z$ and $\nu$. It is possible to prove that this equation can be
satisfied only for $\nu=0$. To investigate the crossover at zero real $\nu$, we
consider purely imaginary frequencies. Multiplying eq.~(\ref{eq:pole}) with
$t_a$ and defining $W = i\nu t_a$, we find
	\begin{align}\label{eq:zeronupole}
	   z = \frac{-2W^2}{e^W-1-W}.
	\end{align}
Stability is achieved when all solutions for $W$ are positive, which is
when $z>-4$.

To find the time dependent variance of $\phi_b$, we evaluate the integral of
eq.~(\ref{eq:defvariance}). Again because of the relative complexity of
$H(\nu)$, the integral cannot be evaluated analytically at arbitrary $t\simeq
t_a$, so we restrict ourselves to $t\gg t_a$. 
For the long timescales ($t\gg t_a$) the integral of eq.~(\ref{eq:defvariance})
is dominated by low frequencies. The integrand is proportional to $1/\nu^2$. The
variance is given by
	\begin{align}\label{eq:defdiffusion}
		\langle[\phi_b(0)-\phi_b(t)]^2 \rangle = 
			\frac{D_\varrho + D_{\tilde v}}
			{(1+z/4)^2} t \equiv D t,
	\end{align}
where we have defined the diffusion constant $D$. The answer is
proportional to diffusion constants in the absence of feedback, where $D_{\tilde
 v} = (e^2/\hbar^2)\langle \tilde v^2 \rangle$ comes from voltage fluctuations
at the superconducting leads and $D_\varrho = (e^2/\hbar^2)G^2 \langle
\varrho^2\rangle $ comes from voltage noise send to the amplifier. We see
that feedback reduces the diffusion constant $D$, the reduction being
proportional to $z^{-2}$ at big $z$. We should mention that $D_\varrho$ grows
with increasing amplification factor $G$, so that the pure increase of $G$ does
not reduce $D$. There is an optimum coefficient which depends on the ratio
between $\langle \tilde v^2 \rangle$ and $\langle \varrho^2\rangle$. In
addition to a term which is linear in $t$, there is also a constant
contribution to the variance that is, in the limit of $t\gg t_a$, given by
	\begin{align}\label{eq:highfreqint}
		&\langle[\phi_b(0)-\phi_b(t)]^2 \rangle_\rs{const}
			\simeq \\
			&[D_\varrho + D_{\tilde v}]\int \left[ \frac{1}{|2i\nu t_a - z\
			H(\nu)|^2} 
			- \frac{1}{(1+z/4)^2\nu^2}     
			 \right] \frac{d\nu}{2\pi} \nonumber \\
			&= [D_\varrho + D_{\tilde v}] t_a\ f(z), \nonumber
	\end{align}
where $f(z)\sim z$ for $z\ll1$ and $f(z)\simeq 1/(2z)$ for $z\gg1$.
We see that also this term is suppressed with increasing $z$, although less
efficient than the diffusion constant. This constant value of the variance is
reached after a typical timescale of $t\simeq t_a/z$.
For $z\gg1$ the high frequency contribution is of the order of $z t_a
D\gg D t_a$. Note that this constant contribution is always larger than $\langle
\tilde v^2 \rangle$, as a consequence of the condition $z\ll \Gamma t_a$. 
Phase coherence is preserved at a timescale $t\simeq t_a$, provided that the
constant part is smaller than $\pi$. 

\section{Extended feedback scheme}
One can extend the feedback scheme to achieve a better reduction of the
diffusion constant and optimization of the constant contribution to the
variance.

The extended feedback scheme is realized with a frequency dependent
amplification factor. The problem we encountered in the previous section is
that the respons time of the circuit, $t\simeq t_a/z$, became smaller with
increasing effective amplification coefficient $z$. Since this response time
should be much larger than the response time of the HJL, $1/\Gamma$, this
restricts the feasible values of $z$. A solution is to increase the
amplification factor at low frequencies, keeping it the same at high
frequencies.

We modify the feedback voltage signal such that its Fourier transform becomes
$V(\nu)\to F(\nu)V(\nu)$, where $F(\nu)$ is the combination of a
PI-filter\cite{regtien} and an extra amplification with a coefficient $r^{-1}$,
where $0<r<1$,
	\begin{align}
	   F(\nu) = \frac{1}{r}\frac{1-i\nu\tau}{1-i\nu\tau/r}.
	\end{align}
With this the amplification coefficient remains unchanged at high
frequencies, $\nu\gg1/\tau$, and is increased with a factor $r^{-1}$ at low
frequencies, $\nu\ll 1/\tau$.
In frequency interval $r/\tau < \nu < 1/\tau$, $F(\nu)$ works as
an integrator of the feedback signal, with $F(\nu) \sim 1/(i\nu\tau)$. 

With this, the expression for the optical phase as reaction on the noises is
modified to
	\begin{align}\label{eq:ftphasesfilter}
	   \frac{\hbar}{2e}\phi_b^\nu &= \frac{-(r-i\nu\tau)\tilde v_\nu + 
			(1-i\nu\tau) G\varrho_\nu}{ 2i\nu t_a(r-i\nu\tau)-  z(1-i\nu\tau)
				H(\nu)}t_a.
	\end{align}
The stability analysis is similar to the case of simple feedback. Instability
would occur at low frequencies and the stability requires $z>-4r$.

We continue with
calculating the variances of the phases. As in the previous section the
diffusion constant is determined by low frequencies and is therefore given by
eq.~\eqref{eq:defdiffusion} with modified value of $z \to z/r$ and diffusion
constant $D_\varrho \to D_\varrho/r^2$,
	\begin{align}
	   D_r = D_{\tilde v} \frac{1+ (z/r)^2 k}{(1+ z/4r)^2},\quad 
		k \equiv \frac{\langle \varrho^2 \rangle}{(\gamma t_a)^2\langle
		\tilde v^2\rangle}
	\end{align}
We rewrite the equation in this form since we would like to optimize it with
respect to the effective feedback coefficient $z/r$. As we mentioned, the
detector noise, $\varrho(t)$, fed to the amplifier, is amplified as well, so
that the optimal value is not infinite. It depends on the effective ratio of
noises $k$ and is given by $z/r=1/4k$, so that the optimal value of the
diffusion constant becomes $D_{r,\rs{opt}}=16 D_{\tilde v}k/(16k+1)$. The
reduction of the diffusion constant is significant at sufficiently small $k$.

It is natural to assume that the integration constant $\tau$ is much bigger
than $t_a$. Then the time dependent fluctuation can be expressed as 
	\begin{align}\label{eq:driftlargetau} 
	   \langle[\phi_b(0)-\phi_b(t)]^2 \rangle = D_r t +(D-D_r) 
			\frac{1-\exp(-\delta t)}{\delta},
	\end{align}
where the high frequency diffusion constant $D$ is given by
eq.~\eqref{eq:defdiffusion}, and $\delta = (4r + z) /[\tau(4+ z)]\simeq
1/\tau$, for $z\gg1$, is the reaction frequency of the circuit. The second term
on the righthand side will become constant in time $t\simeq \tau$. This constant
replaces the constant term of eq.~\eqref{eq:highfreqint}, of the previous case
of simple feedback,
	\begin{align}\label{eq:highfreqintfilter}
		&\langle[\phi_b(0)-\phi_b(t)]^2 \rangle_\rs{const}
			\simeq \frac{D-D_r}{\delta} \\
		& = \frac{16 \tau (D_\varrho+D_{\tilde v})}{(4+z)(4r+z)} -
			\frac{16 \tau (4+z)(r^2+ z^2 k)}{(4r+ z)^3}  D_{\tilde v} \nonumber\\
		& \simeq \frac{(16k)^2\tau}{r(16k+r)(16k+1)}
			\left[ D_\varrho + \frac{D_{\tilde v}}{16k+1} \right], \nonumber
	\end{align}
where in the third line we have reduced the expression by inserting the optimal
value for the feedback $z/r=1/4k$. We find low values for the variance compared
to eq.~\eqref{eq:highfreqint}, when $k\ll1$ and $z\gg1$, such that
$\tau/z^2=\tau/(G \gamma t_a)^2\ll1$.

The diffusion constant of the optical phase, $D_r$, can be thus reduced by
reducing $r$. Let us address the limit of vanishing noise $\varrho$. In this
limit $k\to0$ so that the optimal diffusion constant, $D_{r,\rs{opt}}\simeq16 k
D_{\tilde v}\to 0$. In the 
limit $r\to 0$, the transformation $F(\nu)$ works as a pure integrator for
frequencies $0<\nu < \tau$, with $F(0)\to\infty$. Thereby we achieve a phase
lock with the phase variance given by the limit of
eq.~\eqref{eq:highfreqintfilter} at $r\to0$
	\begin{align}\label{eq:optimalvariance}
		\langle[\phi_b(0)-\phi_b(t)]^2 \rangle_\rs{opt}
			&\simeq \\
			\lim_{r\to0}\frac{D-D_{r,\rs{opt}}}{\delta} &= \frac{4\tau}{z}
				[D_\varrho +D_{\tilde v}]\ll\pi, \nonumber
	\end{align}

\begin{figure}
  \centering
  \onefigure{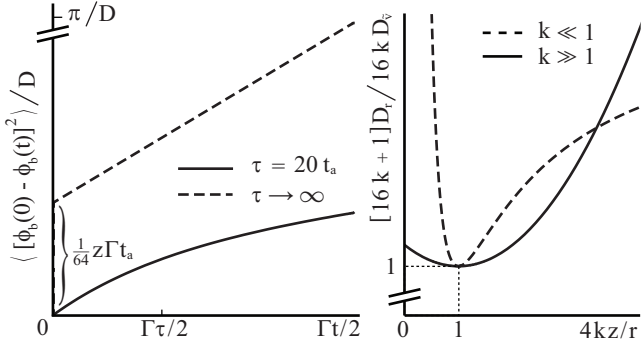}
  \caption{Diffusion in the HJL with feedback. The left panel shows
the time evolution of the variance of $\phi_b$ for two cases. The case of
extended feedback [eq.~\eqref{eq:driftlargetau}, with $\tau=20t_a\gg t_a$] is
represented by the dashed line and the case of simple feedback
[eqs~\eqref{eq:defvariance} and~\eqref{eq:highfreqint}] is represented by the
solid line. The latter can be seen as a special case of the extended feedback
scheme with integration time $\tau\to\infty$. We have taken $z = k/4r\gg1$,
$k\ll r$, $z>\tau$ and $D_r\ll D$. Phase coherence is lost when the variance
reaches a value of the order of $\pi$.
The right panel shows the dependence of the diffusion constant, $D_r$, on $z$,
in two parameter regimes. The optimal value of $z$ is indicated.
  }\label{fig:diffusion}
\end{figure}

To conclude this section, we note that using the extended feedback scheme,
which includes the filter $F(\nu)$, we can significantly reduce both the
diffusion constant and the constant contribution to the variance of the optical
phase of the HJL. The time evolution of the variance of the optical phase in
both the simple and the extended feedback scheme is shown in
fig.~\ref{fig:diffusion}.

\section{Locking two HJLs}
In this section we study an alternative way to stabilize the voltage
fluctuations of a HJL, which is by locking its phase to a stable optical source
of close frequency. This source can be another HJL that is not electrically
connected to the first one, or another laser. We will show that in this case
the feedback quenches the phase diffusion. The fluctuation of the phase remains
just finite at big time differences. If the reference source is ideal, this
provides infinite decoherence time of the HJL. If the reference source is
itself subject to diffusion, the superconducting phase of the HJL follows this
diffusion. Still the fluctuation of the phase difference between the phase of
the HJL and that of the reference source remains finite.

\begin{figure}
  \onefigure{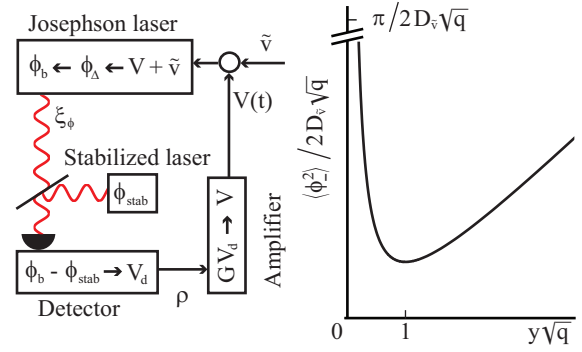}
  \caption{Coupling of two HJLs. The left panel shows the feedback
scheme for the coupling of a HJL to a stable optical source.
The blocks represent elements of the feedback loop. Each element
is characterized by a type of conversion, for instance voltage to voltage
conversion in case of the amplifier. Solid (wavy) lines represent an electrical
(optical) connection between the elements.
The right panel shows the variance of the phase difference,
$\phi_-=\phi_b-\phi_\rs{stab}$, as a function of the feedback strength, $y$,
according to eq.~\eqref{eq:variancephiminus}. This variance is constant over
time. The optimal value is indicated in the plot. 
  }\label{fig:twolasers}
\end{figure}

The locking is modelled using a simple feedback scheme, shown in
fig.~\ref{fig:twolasers}. The scheme is similar
to the one in fig.~\ref{fig:feedback}. The only difference is that the
stabilizer uses as a reference source the external optical signal rather then
the time averaged incoming signal.
This can be achieved by measuring the interference of the light from the
stable optical source with that of the
HJL. The resulting feedback voltage signal becomes $V(t) =
-G [(\hbar\gamma/2e)(\phi_b(t)-\phi_\rs{stab}(t))+\varrho(t)]$, with
$\phi_\rs{stab}(t)$ being the phase of the stable source. Using this,
the Fourier transformed phase, $\phi_b^\nu$, is determined as before [using
eq.~\eqref{eq:lineqsmotion2}]. With this the Fourier transform of the phase
difference, $\phi_- \equiv \phi_b - \phi_\rs{stab}$, satisfies
	\begin{align}\label{eq:twolaserphases}
	   \phi_-  = 
			\frac{\frac{2e}{\hbar}(G\varrho^\nu - \tilde v^\nu)-
			2i\nu \phi_\rs{stab}}{2i\nu - y},
	\end{align} 
where the typical frequency of the feedback circuit is $y\Gamma$, where
dimensionless $y\equiv G\gamma/\Gamma$ must be small to ensure that the
feedback occurs at frequencies smaller than the response frequency of the HJL.
The stability requires $y>0$. 

We note that the locking can occur even if there is a frequency difference
between the stable source and the HJL. We can estimate the frequency difference
at which the HJL remains locked to the stable source from
eq.~\eqref{eq:twolaserphases}. For small frequencies, $\nu\ll y$, we find an
average lag of phase difference proportional to the time derivative of the phase
$\phi_\rs{stab}(t)$, 
	\begin{align}
	   \langle \phi_-(t)\rangle \simeq \frac{2}{y} \frac{d\phi_\rs{stab}(t)}{dt}.
	\end{align}
The lag should much be less than $\pi$ for the feedback to remain in the linear
regime. Therefore, the phase lock persists for frequency differences
$\Delta\omega\ll y\Gamma$.

From eq.~\eqref{eq:twolaserphases} we can infer that the phase difference,
$\phi_-$, is not subjected to diffusion. As shown in previous sections, the
phase of the stabilized HJL diffuses, which is reflected by the pole
at $\nu=0$. Since in the expression for $\phi_-$, the phase $\phi_\rs{stab}$ is
multiplied with $\nu$, there is no longer a pole at $\nu=0$, implying that
$\phi_-$ is not subjected to drift. Hence, the optical phases of the HJLs are
synchronized to each other. 

Let us calculate the variance for the phase difference $\phi_-$. For
this, we use eqs~\eqref{eq:defvariance} and~\eqref{eq:twolaserphases},
assuming $t\to\infty$. Furthermore, we assume an ideal stable reference
source. We find
	\begin{align}\label{eq:variancephiminus}
	   \langle \phi_-^2\rangle = 2\frac{D_\varrho + D_{\tilde v}}{y} =
			2D_{\tilde v}\frac{1 + y^2 q}{y}, 
			\quad q \equiv \frac{\Gamma^2 \langle\varrho^2\rangle} 
				{\gamma^2 \langle\tilde v^2\rangle}.
	\end{align}
In this form, we can easily find the effective feedback
coefficient $y$ that optimizes the variance. It is given by $y=1/\sqrt{q}$.
For the phase lock to persist, the variance should be much less than $\pi$. 
Figure~\ref{fig:twolasers} contains a plot of the variance as a function of $y$,
with the optimal value of $y$ indicated.

The locking described, yields a new way to control the superconducting phase
difference. The optical phase can be easily changed by changing the optical
path lengths of the laser beams. This change can be incorporated into the
change of delay times $t_d$ of the beams, $\delta\phi_\rs{stab} = (\omega_J/2)
\delta t_d$. Because of the phase lock $\Phi_b$ changes accordingly and so does
the superconducting phase difference between the leads. Owing to the Josephson
relation, one can produce voltage pulses by changing the optical path length in
time.

\section{Conclusions}
In this paper, we propose stabilization of voltage fluctuations in a
half-Josephson laser (HJL), by means of optical feedback. Using a feedback
scheme, based on the well-known {\it Pound-Drever-Hall} stabilizer, we can
significantly decrease the diffusion constant of the phase. The feedback can be
further enhanced using the frequency dependent amplification. 
In the second feedback scheme, the voltage fluctuations of the HJL are
stabilized by locking it to a stable optical source. The variation of the phase
difference does not grow with increasing time. We have shown that with this
feedback scheme one can achieve the control of superconducting phase difference
by changing optical path lengths. 
These proposals prove the application potential of the HJL and demonstrate
this to be an interesting tool to combine superconductivity and optics.

\acknowledgments
We acknowledge financial support from the Dutch Science Foundation NWO/FOM.

\end{document}